\documentclass[twocolumn]{revtex4}

\usepackage{graphicx}

\newcommand{\D}{\mathrm{d}}

\newcommand{\kb}{k_{\mathrm{B}}}
\newcommand{\kbt}{k_{\mathrm{B}}T}

\newcommand{\lb}{l_\mathrm{B}}
\newcommand{\ld}{\lambda_\mathrm{D}}
\newcommand{\lgc}{\lambda_\mathrm{GC}}

\newcommand{\epsz}{\varepsilon_{0}}

\newcommand{\epsn}{\varepsilon(n)}
\newcommand{\epss}{\varepsilon_s}

\begin{document}


\title{
Dielectric decrement as a source of ion specific effects}
\author{Dan Ben-Yaakov, David Andelman}
\email{andelman@post.tau.ac.il}
\affiliation{Raymond and Beverly Sackler School of Physics and Astronomy\\ Tel Aviv
University, Ramat Aviv, Tel Aviv 69978, Israel}

\author{Rudi Podgornik}
\affiliation{Department of Theoretical Physics\\ J. Stefan
Institute, Department of Physics\\ Faculty of Mathematics and Physics and
Institute of Biophysics\\ Medical Faculty, University of Ljubljana\\ 1000 Ljubljana, Slovenia}

\date{11/12/2010}

\begin{abstract}
Many theoretical studies were devoted in the past to ion-specific effects, trying to interpret a large body of experimental evidence, such as surface tension at air/water interfaces and force measurements between charged objects. Although several mechanisms were suggested to explain the results, such as dispersion forces and specific surface-ion interactions, we would like to suggest another source of ion-specificity, originating from the local variations of the dielectric constant due to the presence of ions in the solution. We present a mean-field model to account for the heterogeneity of the dielectric constant caused by the ions. In particular, for ions that decrease the dielectric constant we find a depletion of ions  from the vicinity of  charged surfaces. For a two-plate system, the same effect leads to an increase of the pressure in between two surfaces. Our results suggest that the effect of ions on the local dielectric constant should be taken into account when interpreting experiments that address ion-specific effects.
\end{abstract}

\maketitle
\section{introduction}

The first interest in ion specific effects dates back to the end of the 19th century, when Franz Hofmeister and his co-workers~\cite{hofmeister} measured the thermodynamic properties of protein precipitation in various salt solutions. They tested numerous  salt species and found that the protein solubility properties can be arranged in a specific ionic order. This classification of cations and anions is known today as the {\it Hofmeister series}. Since then many other experiments demonstrated that ion-specific effects take place in a large variety of chemical and biological systems~\cite{collins1985}. These experiments include measurement of surface tension at air/water interface \cite{heydweiller1910,air_water_2,air_water_3}, proteins stability \cite{collins2004}, forces between charged surfaces such as mica or silica surfaces~\cite{dishon,pashely}, osmotic pressure in systems containing biological macromolecules \cite{parsegian1992,podgornik1994,parsegian2005} and many more.

Noticeable efforts have been devoted over the years to understand the physical mechanisms that lead to ion-specific effects~\cite{collins1985,ruckenstein2003a,ruckenstein2003b,ninham2004,kunz2010}. A typical problem is to obtain density profiles of ions near charged surfaces such as macromolecules, membranes and colloids or neutral dielectric
interfaces ({\it e.g.,} air/water interface), and use the profiles to calculate macroscopic quantities such as osmotic pressure and surface tension.

Theoretical endeavors offer a variety of perspectives regarding the origin of the ion-specific effects. Most recently, dispersion interactions~\cite{ninham1997,ninham2004} depending on the polarizability of the ions, have been proposed to furnish the missing link between ionic profiles and ion-specific interactions. These interactions add an additional term to the total ionic energy close to a non-charged dielectric interface. This term varies as $B/z^3$, where $z$ is the distance of the ion from the surface, located at $z=0$. The ion-specificity then emerges from a variation in the specific ion polarizability, leading to different values of the coefficient $B$.

Apart from the dispersion interactions, ion-hydration interaction~\cite{ruckenstein2003a,ruckenstein2003b} has been
suggested to lead to an effective short-range ion-specific interaction with the bounding surface. For example, an attractive potential of the ions to the surface in the shape of a square well, takes into account their larger affinity as compared to ion-water interaction. This attraction then boosts the concentration of the ions in the surface vicinity.

In a similar fashion but with an opposite effect, Onsager and Samaras dealt in their seminal work~\cite{onsager_samaras} with repulsive image charge interactions as a source of ionic depletion from a non-charged dielectric interface. Image charge interaction is a short-range surface interaction, having an $\exp(-2z/\ld)/z$ dependence, where $\ld$ is the Debye screening length. This repulsive interaction becomes significant for air/water interfaces and is responsible for the increase of surface tension of saline solutions as compared with pure water. More recently,  additional interactions have been proposed to augment the original Onsager and Samaras model, and offer an explanation to ion-specific surface tension of electrolyte solutions, in agreement with experiments~\cite{levin200x,levin200y}.

In another approach~\cite{collins2004} it was suggested that surface charge density is a crucial parameter controlling ionic specificity. The density is determined by the net charge and specific volume (i.e., radii) of the ions. For small ions, the surface charge density can be large, leading to a strong attractive interaction with the polar water molecules. This results in a ``hard" hydration shell with high energy cost of removing it. Oppositely, for large ions, the surface charge density is smaller, and therefore, the energy cost of breaking the ion-water complex is lower, leading to a ``soft" hydration shell. Due to the different energy costs of ``hard" and ``soft" hydration shells, it is favorable to have pairing where both cations and anions are either small or large. An interesting feature of this ionic pairing mechanism is a possible explanation of a reversal in the order of the Hofmeister series, as observed in several experiments~\cite{collins2004}.

Besides the analytical approaches, computer simulations gain much needed insight into the intricacies of the ionic specificity~\cite{netz,jungwirth}, and are used to investigate various cross-interactions between the system components (solutes, solvent and surfaces). Computer simulations capture a very detailed microscopic picture as compared to continuum theories such as the Poisson-Boltzmann (PB) theory and its extensions. However, they often lack the simplicity and predictive power of analytical models and their straightforward intuition.

From the current state of theoretical and experimental studies, it appears to be quite clear that ion-specific effects are due to the interplay between the ion-ion, ion-solvent and ion-surface interactions, taking into consideration both electrostatic and non-electrostatic interactions. However, it is still not well understood what are the important physical and chemical properties of the solutes and solvent that determine these specific interactions.

The traditional PB theory accounts for electrostatic interactions on a mean-field level with well understood drawbacks and limitations \cite{Ali}.  It neglects all ion-specific effects except for the ion valency, which means that it treats identically all ions of the same valency. Apart from the electrostatic interactions, other non-electrostatic interactions can be added into the theory. This was done in the standard DLVO theory of colloid stability~\cite{verwey1948}, where the total interaction decomposes into a sum of the van der Waals and electrostatic interactions.

In the present work we would like to suggest a phenomenological approach to treat different ionic species. Our approach shares some of the simplicity of the PB theory, while  taking into account the ionic specificity on a mean-field level. In particular, we focus on the effect the ions have on the local solution polarity as codified by its static dielectric function. The static dielectric function is treated as a spatially-dependent function $\varepsilon(z)$, where the spatial dependence is given implicitly by the local ionic concentration,  thus introducing ionic specificity in an implicit manner. In the next section we elaborate on the dependence of the dielectric constant on the ionic concentration and discuss its physical range. In section~\ref{model}, we present a mean-field model that accounts for the effect of the ions on the dielectric constants. Finally, results for density profiles and the inter-plate pressure are presented in sections~\ref{profile} and \ref{pressure}, respectively.

\section{Ion dependent dielectric constant}\label{dielectric}
\begin{table}
\caption{\textsf{Values of the linear coefficient $\beta$, Eq.~(\ref{phenom}), in units of M$^{-1}$ for various salts. [adopted from the references as indicated].}}
\begin{tabular}{| c | c | c | c | c | c |}
  \hline
  & Ref.~\cite{hasted1973} & Ref.~\cite{glueck1964} &
  Ref.~\cite{sirdhar1990,sirdhar1991} & Ref.~\cite{may1999} & Ref.~\cite{chandra2000}\\ \hline
  HCl & -20 &  -18.1 & -- & -- & --\\
  LiCl & -14 & -12.6 & -11.55 & -- & -- \\
  NaCl & -11 & -11.8 & -- & -12.8 & -10.77 \\
  KCl & -10 & -11.8 & -- & -- & -8.27 \\
  CsCl & -- & -12.6 & -7.79 & -- & --\\
  RbCl & -10 & -- & -7.96 & -- & \\
  NaF & -12 & -11.8 & -- & -- & -- \\
  KF & -13 & -11 & -- & -- & --\\
  CsF & -- & -12.6 & -- & -- & --\\
  LiI & -- & -14.95 & -- & -- & -- \\
  NaI & -15 & -14.17 & -- & -- & -- \\
  KI & -16 & -13.38 & -- & -- & -- \\
  CsI & -- & -14.17 & -- & -- & -- \\
  NaOH & -21 & -- & -- & -- & -- \\
  \hline
\end{tabular}
\end{table}

The static dielectric function of an electrolyte solution is generally found to be smaller than that of the pure solvent \cite{debye,hasted1973}. This decrement of the dielectric constant has been attributed to various sources, which underly the changes in the dielectric response of the solution. We mention here two of the most important ones:

{\textbf{\textit{Ionic polarizability.}}}~~Each ion in an aqueous solution creates a cavity and displaces one or several water molecules in this process. Since the ions usually have smaller static and dynamic polarizabilities than water, they modify the dielectric response of the solution~\cite{paunovic}. This effect, though omnipresent, is nevertheless small and does not have substantial consequences.

\begin{figure}
\includegraphics[width=0.48\textwidth]{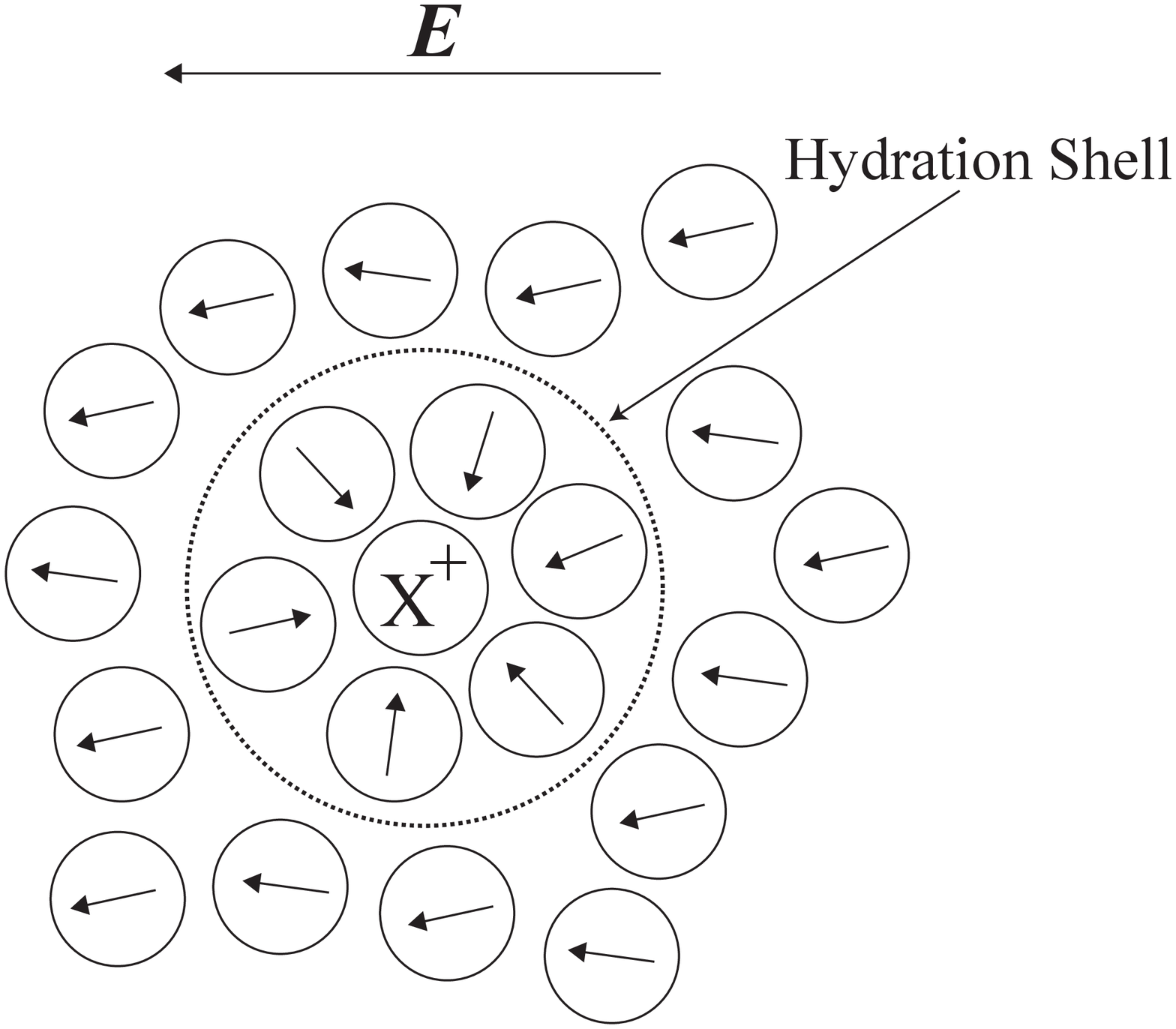}
\caption{\textsf{A schematic drawing of water molecules around a monovalent cation in presence of an electrostatic field, $\mathbf{E}$. Arrows represent the dipoles of water molecules while the hydrogen bond network is not shown explicitly. The water molecules in the vicinity of the cation form a hydration shell and are oriented along the field lines generated by the cation, leading to a decrease of their contribution to the screening of the external electrostatic field. The more remote molecules are less affected by the cation and orient themselves along the external field lines. Furthermore, the cation polarizability results in a different contribution to the electrostatic screening  as compared to the contribution of a water molecule.}\label{fig1}}
\end{figure}

\textbf{\textit{Hydration shell.}}~~The gist of the dielectric decrement in aqueous electrolytes is connected with the structural modification of water molecules in the immediate proximity to the ion due to large electrostatic fields emanating from the dissolved ions~\cite{paunovic}. By the action of these strong electrostatic fields, hydration shells are formed around solvated ions, where nearby water molecules are oriented along the ion electrostatic field, leading to an additional pronounced dielectric decrement as is depicted schematically in Fig.~\ref{fig1}.

The lowering of the dielectric constant in the vicinity of an ion can be attributed to the effects described above, which represent the quintessential mechanisms of the dielectric decrement. In an ionic solution the same effect would lead to an overall reduction of the total solution dielectric constant as a function of the ionic concentration. In addition to these primary mechanisms, there are other possible sources of the dielectric decrement such as excluded volume~\cite{carnie1988} and dynamical effects~\cite{onsager1977} that can have a noticeable contribution.

In numerous experimental studies~\cite{glueck1964,sirdhar1990,sirdhar1991,may1999}
and simulations~\cite{chandra2000}, the dependence of the dielectric constant was found to depend linearly on the salt concentration $n$ for molar concentrations ranging between zero and 1.5\,M:

\begin{equation} \label{phenom} \epsn=\epsz+\beta n\, ,
\end{equation}
where $\epsn$ is the dielectric constant of the ionic solution, $\epsz$ is the dielectric constant of pure solvent, and $\beta$ is a phenomenological coefficient (in units of M$^{-1}$) of the linear term. The most common case where $\beta<0$  corresponds to a dielectric decrement. Throughout this paper the value of $\epsz=80$ is taken to be that of pure water.

The dielectric decrement parameter $\beta$ has salt-specific values~\cite{hasted1973,glueck1964,sirdhar1990,sirdhar1991,may1999,chandra2000}. In Table~1 we show a range of such values adopted from different experiments and computer simulations. The cited values are all negative and vary from about -8 to -21 (dielectric decrement). There is also a noticeable spread of the $\beta$ values, especially when one compares experiments with computer simulations. Examining the $\beta$ values given in Refs.~\cite{hasted1973} and \cite{glueck1964} for homologous series of the halides: Cl, F and I, and the alkaline metals: Li, Na, K, Cs, and Rb, some remarks can be made.  In some cases, a  trend can be seen, where the magnitude of $\beta$ decreases as the ionic radius (size) increases for the alkalines. This can be  seen for the homologous series of XCl where X is the alkaline ion. For the homologous halide series, the trend of $|\beta|$ with the halide ionic size is not monotonous. We note that the above linear form of the static dielectric constant $\varepsilon(n)$ depends on the concentration of the salt and cannot, in general, be entangled in a simple manner into separate ionic contributions of the cations and anions. Any such separation requires additional hypotheses and experiments that are unavailable at present.

\section{The model}\label{model}

In what follows we concentrate on the dielectric decrement effects of ions (when $\beta$ is negative), and systematically assess the modification wrought in the equilibrium ionic profiles as well as interactions between two apposed charged surfaces. Note that we treat the problem of counter-ions only (no co-ions) in order to simplify the model and results.

We set up a phenomenological model based on a modification of the standard PB theory, which includes the effect of the counter-ion concentration $n(\vec{r})$ on the local dielectric constant, in a consistent manner, via the local variation of $n(\vec{r})$ in $\varepsilon(n)$. This model can be formulated in two distinct but equivalent versions. First, one can dress it in the form of the modified PB theory starting with the appropriate generalization of the PB free-energy functional where the fundamental quantity is the local electrostatic potential. Equivalently, as is shown in the Appendix, the PB free-energy functional can also be cast  in the form of a density functional theory (DFT), where the fundamental quantity is the local density of the counter-ions, $n(\vec{r})$.

The mean-field free energy is similar to the regular PB {\sl ansatz}, except for the coupling between the electrostatic field,
$-\nabla\psi$, and the positively charged counter-ion concentration, $n$. For simplicity, we delimit ourselves to systems that are
translationally invariant in the lateral ($x,y$) directions having a spatial dependence only on the $z$-direction. The free energy then reads:
\begin{eqnarray}
\label{free_energy}& F[\psi', \psi, n]/{\cal A}=\int\D
z\left[-\frac{\varepsilon(n)}{8\pi}\left(\psi'\right)^2 +~ e n \psi \right.  \nonumber\\
 &   \left.  +~\kbt n(\log n
-1)-\mu n\right]+\sum_{s}e|\sigma|\psi_s\, ,
\end{eqnarray}
%
where ${\cal A}$ is the lateral cross-sectional area, $e$ is the electron charge, $\kb$ the Boltzmann constant, $T$ the temperature and $\mu$ the chemical potential. The last term is the contribution of the charged surfaces to the free energy, where $-\sigma<0$ is the {\it negative} surface charge density,
$\psi_s$  is the electrostatic surface potential, and the last term is summed over all charged surfaces.  Note that throughout the manuscript  $e\sigma$ is the surface charge density and $\sigma$ is the corresponding number density.

Using the variational principle for the free energy, Eq.~(\ref{free_energy}), with respect to $n$ and $\psi$ gives the Euler--Lagrange equations determining the electrostatic potential and density profiles:
\begin{eqnarray}
\label{eqlbrm_eq_1}
\frac{\delta F}{\delta \psi}=0~ &\Rightarrow &~ 4\pi en +\frac{\D}{\D z}\bigg(\varepsilon(n) \psi'\bigg)=0 \, ,\\
\label{eqlbrm_eq_2} \frac{\delta F}{\delta
n}=0 ~ &\Rightarrow &~ \mu=-\frac{1}{8\pi}\varepsilon'(n) \psi'^2
+e\psi+\kbt\log n \, ,\nonumber\\
& &
\end{eqnarray}
where $\varepsilon'(n)=\D\varepsilon/\D n$.

As was discussed in our previous work~\cite{benyaakov2009}, as long as the system has only one-dimensional spatial inhomogeneity, the pressure is a $z$-independent constant, and in fact, represents the first integral of the Euler--Lagrange equations. Using Eq.~(25) of Ref.~\cite{benyaakov2009}, we arrive at the following form of the pressure, ${P}$

\begin{equation} \label{pressure1} {P}=-\frac{1}{8\pi}\left[\varepsilon(n)+
\varepsilon'(n) n\right]\psi'^2+ \kbt n = {\rm const}\,.
\end{equation}
The pressure is composed of an appropriately modified Maxwell stress tensor, which takes into account the density dependence of the dielectric constant \cite{landau}, as well as the standard van't Hoff term proportional to the counter-ion concentration.

Combining Eqs.~(\ref{eqlbrm_eq_1})-(\ref{pressure1}), the following first-order differential equation for $n$ is derived:
\begin{equation}\label{n_ode}
\frac{\D n}{\D z}=-\sqrt{\frac{2\pi e^2}{\kbt}}\frac{n}{f'(n)}\quad ,
\end{equation}
where
\begin{equation}\label{f_n}
f(n)=\varepsilon(n)\sqrt{\frac{n-\hat{P}}{\varepsilon(n)+n\varepsilon'(n)} }\quad ,
\end{equation}
and $\hat{P}=P/\kbt$.
The solution of Eq.~(\ref{n_ode}) yields the equilibrium profile of the counter-ion density. The boundary condition at the charged interface is obtained by taking the variation of Eq.~(\ref{free_energy}) with respect to the surface potential $\psi_s$:
\begin{equation}
\label{bc}
\varepsilon_{s} {\frac{\D\psi}{\D z}}\bigg|_{s}=4\pi e\sigma\, ,
\end{equation}
where $\varepsilon_s=\varepsilon(n_s)$ is the value of the dielectric constant extrapolated to the surface, and $n_s$ is the counter-ion concentration at the surface.

\section{The Single Plate Case} \label{profile}

We first consider the case of a single charged plate placed at $z=0$. This case can be regarded as a two-plate system in the limit of infinite inter-plate spacing, where the pressure in Eq.~(\ref{pressure1}) vanishes, ${P}=0$. The function $f(n)$ from Eq.~(\ref{f_n}) is then given by:
\begin{equation}
f(n)=\varepsilon(n)\sqrt{\frac{n}{\varepsilon(n)+n\varepsilon'(n)} }\quad .\label{f_n_1S}
\end{equation}
Evaluating Eq.~(\ref{pressure1}) at the surface with $\varepsilon=\varepsilon_s$ and $n=n_s$, we obtain an algebraic equation for $n_s$, given by:
\begin{equation} \label{bc_general}
n_{s}=\frac{2\pi e^2 \sigma^2}{\kbt\varepsilon^2(n_{s})}
\left(\varepsilon(n_{s})+\frac{\D \varepsilon}{\D
n}\bigg|_{s} n_{s}\right)\,.
\end{equation}

As has already been noted in the previous section, for ionic concentrations of $0<n< 1.5\,$M, the function $\epsn$ can be regarded to a reasonable accuracy as linear in $n$, Eq.~(\ref{phenom}). This is a key assumption of our model where all the effects discussed in Sec.~\ref{dielectric} contribute to a single phenomenological parameter $\beta=\D \varepsilon / \D n$.  We stress that the linear dependence mentioned in the previous sentence is a phenomenological approach and cannot be justified theoretically for any value of concentration \cite{phenom}.

The derivative of $f(n)$ is given by:
\begin{equation}
 f'(n)=\frac{2 \beta^2
n^2+(\epsz+\beta n)(\epsz+2\beta
n)}{2\sqrt{n(\epsz+2\beta n)^3}}\,
,\label{f_n_linear}
\end{equation}
and the boundary condition, Eq.~(\ref{bc_general}), can be expressed as a cubic polynomial in $n_s$:

\begin{equation}
\beta^2 n_s^3+2\beta \varepsilon_0 n_s^2+\left(\varepsilon_0^2-\frac{4\pi
e^2\sigma^2\beta}{\kbt}\right)n_s-\frac{2\pi
e^2 \sigma^2 \varepsilon_0}{\kbt}=0\,
.\label{bc_1S}
\end{equation}

\subsection{The $\beta=0$ profile}

The equation for the counter-ion density profile that we derived above, Eq. (\ref{n_ode}), appears very different from what one finds in the standard PB theory.
Let us show that the PB solution is indeed obtained by setting $\beta=0$ in our model. Substituting $\beta = 0$ in Eq.~(\ref{f_n_linear}), the density profile obeys the following equation:
\begin{equation}
\label{rudi2} \frac{\D n_{\rm PB}}{\D z}=-\sqrt{\frac{8\pi e^2}{\kbt
\varepsilon_0}} n_{\rm PB}^{3/2},
\end{equation}
and its solution reproduces exactly the familiar profile of Gouy and Chapman (GC)~\cite{andelman2006}:
\begin{equation}
 \label{n_gc} n_{\rm PB}(z)=\frac{1}{2\pi\lb}\left(\frac{1}{z+\lgc}\right)^2\, ,
\end{equation}
where $\lgc=1/(2\pi\lb\sigma)$ is the GC length, $\lb=e^2/(\epsz\kbt)$ is the Bjerrum length and $\epss=\epsz$ for the standard PB case. The counter-ion concentration at the surface is obtained via the Grahame equation, $n^{\rm PB}_s=2\pi \lb \sigma^2$ \cite{andelman2006}, or alternatively can be expressed in terms of $\lgc$, $n_s^{\rm PB}=1/(2\pi\lb\lgc^2)$.

\begin{figure*}
\includegraphics[width=0.8\textwidth]{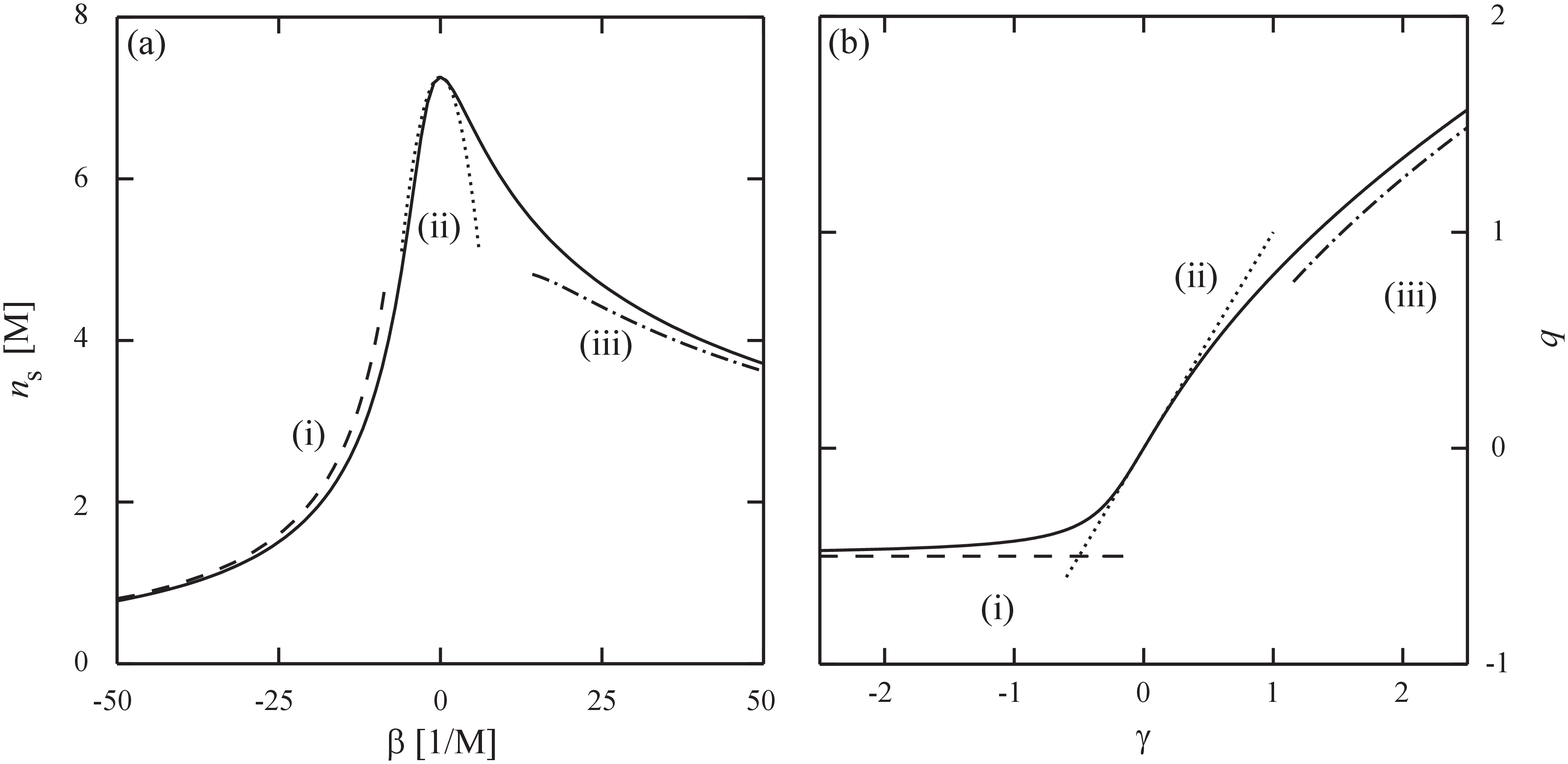}
\caption{\textsf{In (a) the ion concentration at the interface, $n_s$, is plotted as a function of $\beta$. The inverse of Eq.~(\ref{gamma_x}) is denoted by the solid line. The surface charge density is $\sigma=0.01$\,\AA$^{-2}$. Approximated analytical solutions for several limits are denoted by roman numerals (see text). In (b) the parameter $q$ is plotted as a function of $\gamma$. The inverse of the exact relation, Eq.~(\ref{gamma_q}), is denoted by the solid line. The approximated analytical solutions are denoted by the same notations as in (a).}} \label{fig2}
\end{figure*}

\subsection{The boundary condition for $\beta \neq 0$}

Returning to the general $\beta\neq 0$ case, the counter-ion concentration adjacent to the surface, $n_s$, satisfies the more general relation given in Eq.~(\ref{bc_1S}). Introducing the dimensionless parameters $\gamma\equiv \beta n_s^{\rm PB}/\varepsilon_0$ and $\alpha\equiv n_s/n_s^{\rm PB}$,  Eq.~(\ref{bc_1S}) can be rewritten as:
\begin{equation}
\gamma=\frac{1-\alpha\pm\sqrt{1-\alpha}}{\alpha^2}\, .
\label{gamma_x}
\end{equation}
where $n_s^{\rm PB}\sim\sigma^2$ as obtained from the above mentioned Grahame equation.

It is evident from Eq.~(\ref{gamma_x}) that for any $\beta \neq 0$, $\alpha=n_s/n_s^{\rm PB}<1$ and the counter-ion close to the surface, is reduced with respect to PB value. Surprisingly, even when $\beta>0$ and the local dielectric constant increases (electrostatically favorable), there is an effective repulsion of the ions from the interface.

It is also instructive to express Eq.~(\ref{gamma_x}) in terms of the dimensionless parameter $q=\varepsilon_s/\varepsilon_0-1=\beta n_s/\varepsilon_0$, which measures the $\beta$-dependent relative change of the  dielectric constant at the interface. The dimensionless algebraic equation of the boundary condition, Eq.~(\ref{gamma_x}), can now be written in terms of $\gamma$ and $q$,
\begin{equation}
q^3+2q^2+(1-2\gamma)q-\gamma=0\, ,
\end{equation}
and its solution yields:
\begin{equation}
\gamma=\frac{q^3+2q^2+q}{1+2q}\, .
\label{gamma_q}
\end{equation}
Using Eq.~(\ref{gamma_q}) one can analyze the dependence of $\varepsilon_s$, the dielectric constant value at the interface, on $\beta$.

\begin{figure*}
\includegraphics[width=0.4\textwidth]{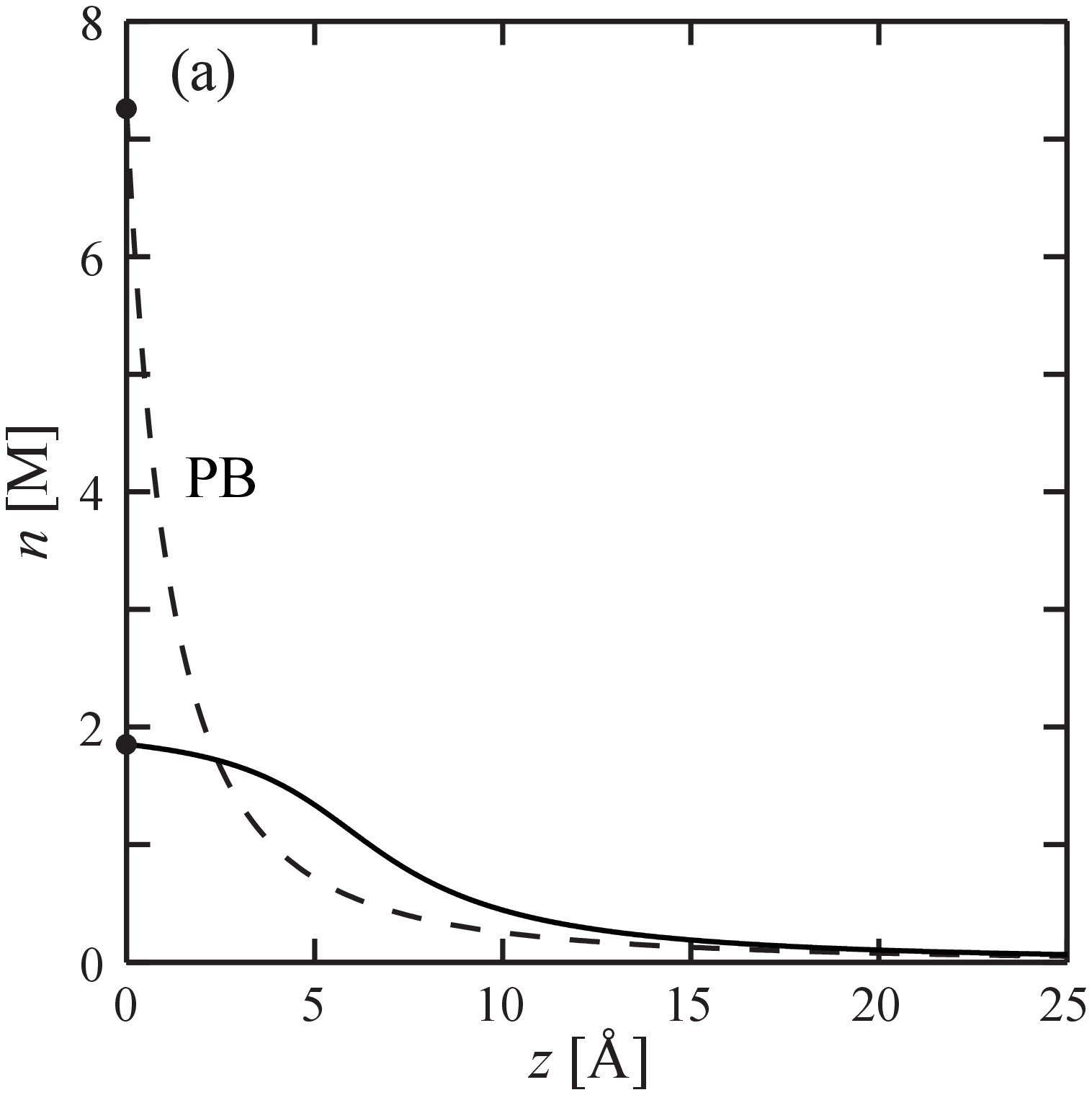}
\includegraphics[width=0.4\textwidth]{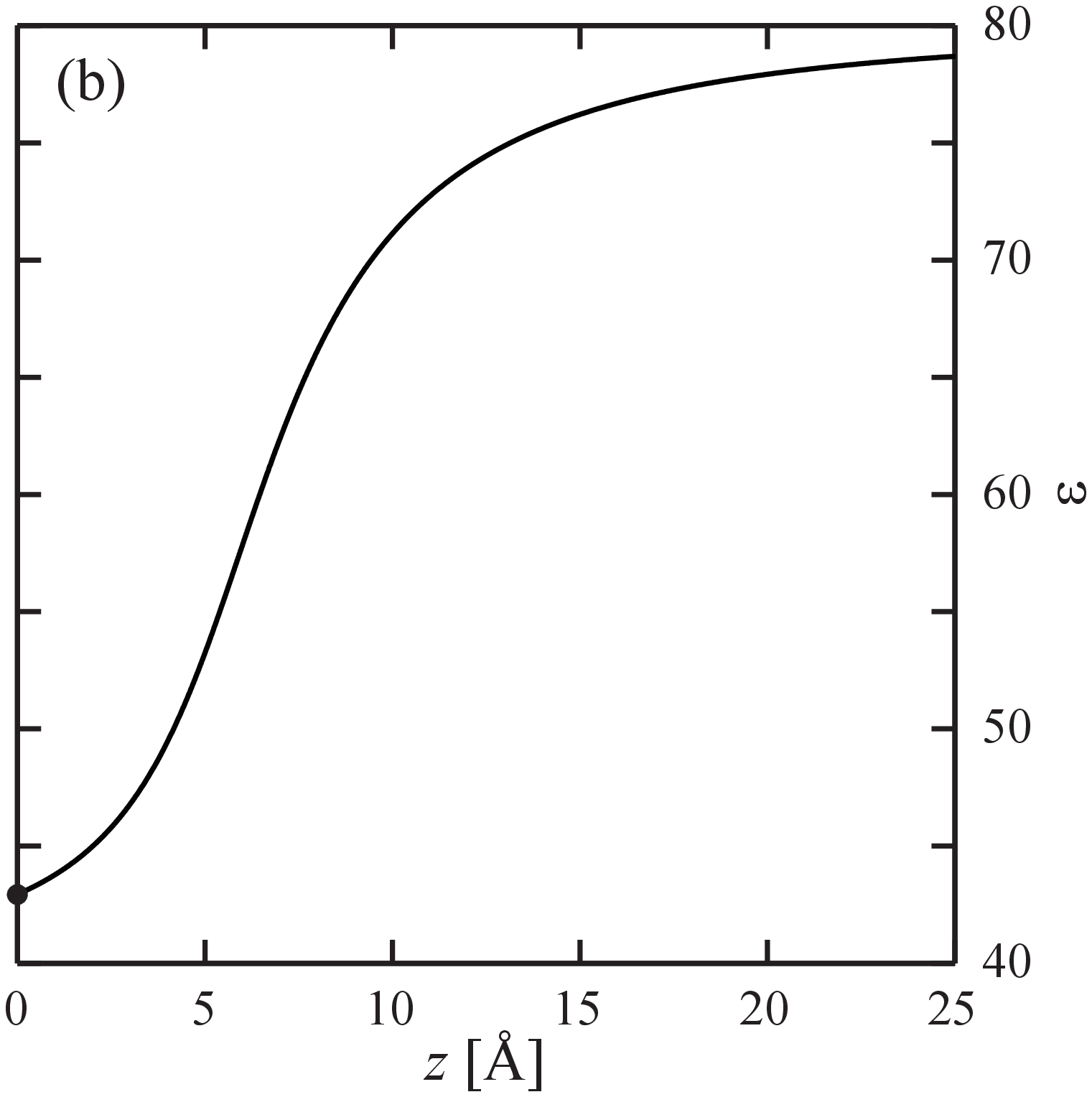}
\caption{\textsf{Counter-ion concentration $n(z)$ and local dielectric constant $\varepsilon(z)$ profiles for $\gamma=-1.8$. In (a) the solid line shows the counter-ion concentration for negative $\beta=-20\,$M$^{-1}$. The dashed line corresponds to
the standard PB case ($\beta=0$), as given by Eq.~(\ref{n_gc}). In (b) the local dielectric constant $\varepsilon(z)$ is shown for the same $\beta$ value. The surface charge density is $\sigma=0.01\,$\AA$^{-2}$ and the calculated surface values are $n_s^{\rm PB}=7.3$\,M, $n_s=1.9$\,M and $\epss=43$.
}} \label{fig3}
\end{figure*}

In several limits the expressions of $n_s(\beta)$ and $q(\gamma)$ simplify:
\\
(\textit{i})
For strongly negative $\beta\ll-1$, equivalent to  $\gamma \ll -1$, the counter-ion concentration at the surface is decaying:
\begin{equation}
n_s\simeq-\frac{{\varepsilon_0}}{{2\beta }}\sim-\beta^{-1}\, ,
\label{bc_1S_large_beta_minus}
\end{equation}
while the change in the surface dielectric constant approaches a limiting value $q=-0.5$, leading to $\varepsilon_s=\varepsilon_0 /2$. For these values of $n_s$ and $\epss$ the polarization energy [see right hand side of Eq.~(\ref{pressure1})] associated with the water molecules is equal the energy of the ions ``effective" polarization, yielding a limiting value for the electrostatic energy.
\\
(\textit{ii})
For small $|\beta| \ll 1$ or $|\gamma|\ll 1$, the correction to $n_s$ is small and given by:
\begin{equation}
\frac{n_s-n_s^{\rm PB}}{n_s^{\rm PB}}= -\left(\frac{\beta n_s^{\rm PB}}{\varepsilon_0}\right)^2\, ,
\label{bc_1S_small_beta}
\end{equation}
while the leading order of $q$  is linear in $\gamma$, $q(\gamma)\simeq\gamma$.
\\
(\textit{iii}) In the limit of extremely polarizable ions ($\beta,\gamma\gg1$), $n_s$ varies as:
\begin{equation}
 n_s\simeq\sqrt{\frac{2\varepsilon_0 n_s^{\rm PB}}{\beta}}\left[1-\frac{3}{4}\sqrt{\frac{\varepsilon_0}{2\beta n_s^{\rm PB}}}\,\right]\sim\beta^{-1/2}\, ,
\end{equation}
and $q$ (or $\varepsilon_s$) grows as
\begin{equation}
q=\sqrt{2\gamma}\left(1-\frac{3}{4\sqrt{2\gamma}}\right)\sim\sqrt{\gamma}\, .
\end{equation}
One would expect that for $\beta>0$, where the ions increase the dielectric constant, the surface ionic density $n_s$ will be increased as well. Surprisingly, $n_s$ is decreased for increased $\beta$ as a result of the interplay between all the contributions, see \cite{pos_beta}.

The exact dependence of $n_s(\beta)$ from Eq.~(\ref{gamma_x}), and $q(\gamma)$ from Eq.~(\ref{gamma_q}), is compared with the above limiting expressions  in Fig.~\ref{fig2}. There is a smooth crossover at $\gamma_{12}\simeq -1$ between  regimes (\textit{i}) and (\textit{ii}), and similarly at $\gamma_{23}\simeq1$ for the crossover between regimes (\textit{ii}) and (\textit{iii}). The crossover values are taken to satisfy the conditions of the approximations. Calculating higher order terms for the three regimes gives intersection values that are similar to our crossover values.

\begin{figure*}
\includegraphics[width=0.4\textwidth]{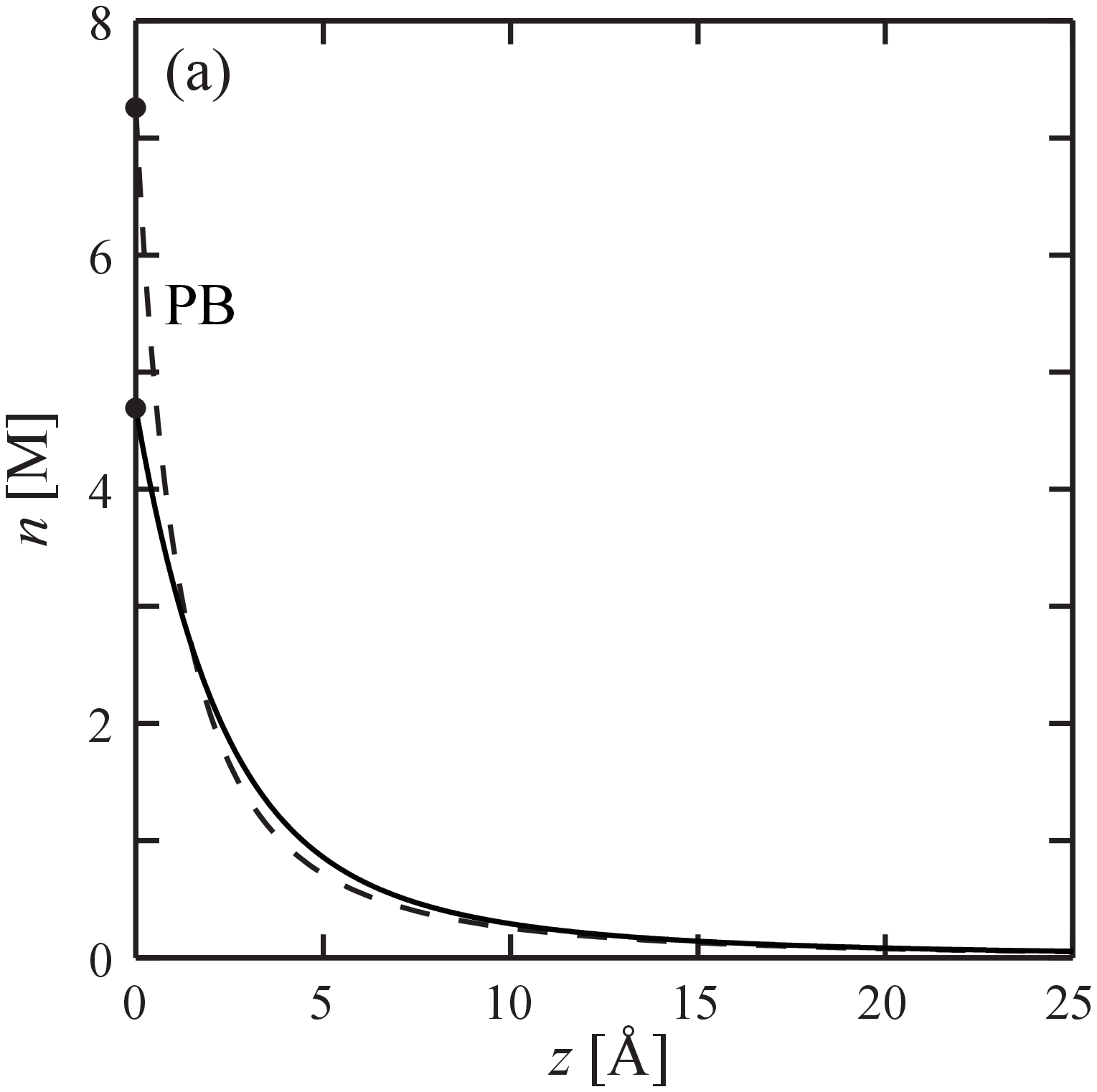}
\includegraphics[width=0.4\textwidth]{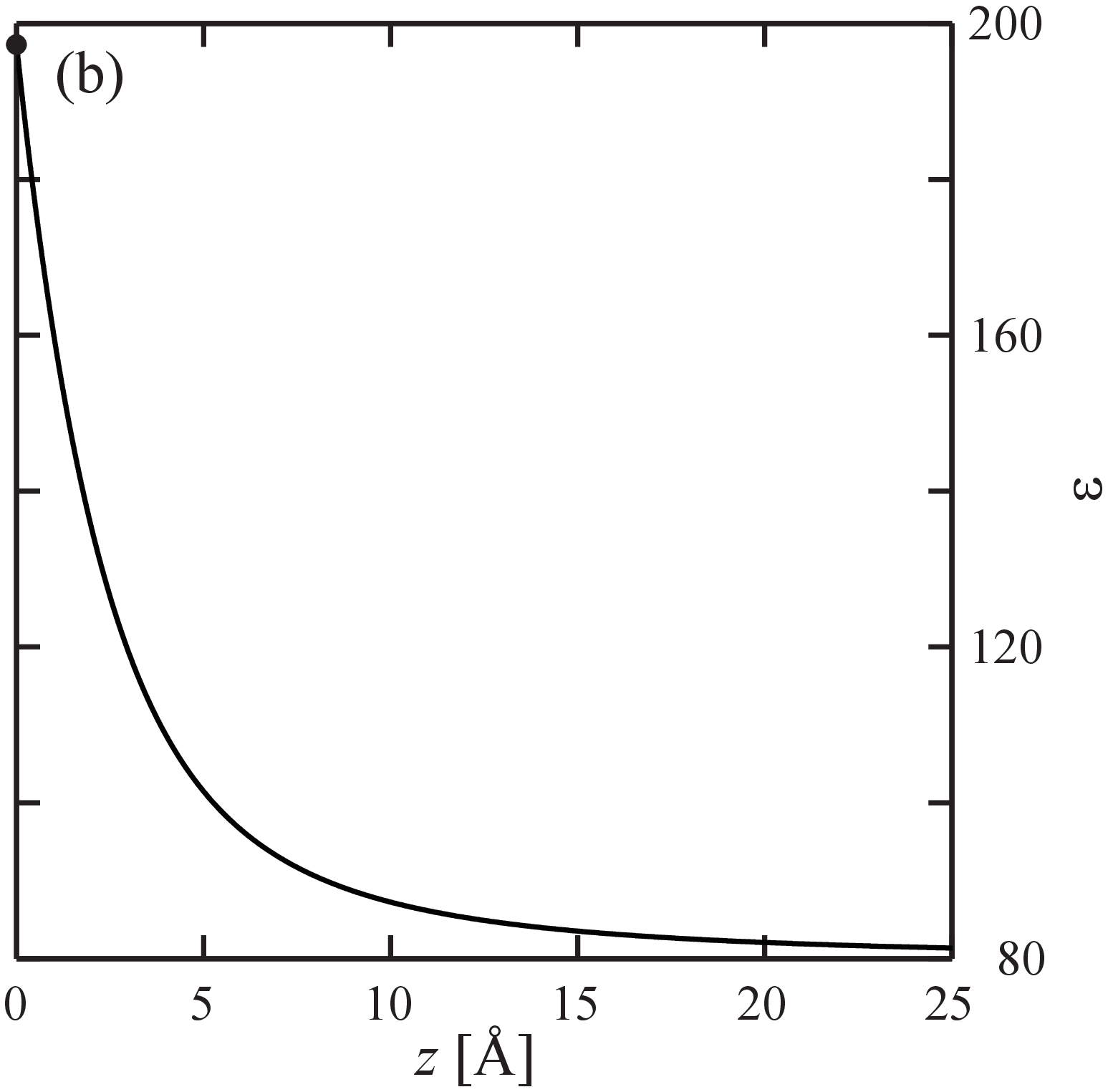}
\caption{\textsf{Counter-ion concentration profile $n(z)$ and local dielectric constant $\varepsilon(z)$ for $\gamma=2.3>\gamma_{23}$. In (a) the solid line shows the counter-ion concentration for positive $\beta=25\,$M$^{-1}$.  The dashed line corresponds to the standard PB case ($\beta=0$), as given by Eq.~(\ref{n_gc}). In (b) the corresponding local dielectric constant is shown for the same $\beta$ value. The surface charge density is $\sigma=0.01\,$\AA$^{-2}$ and the calculated surface values are $n_s^{\rm PB}=7.3$\,M, $n_s=4.7$\,M and $\epss=197$.
}} \label{fig4}
\end{figure*}

\subsection{The profiles for $\beta\neq 0$} \label{sec_profile_beta}

The above analysis for the limiting regimes for $q(\gamma)$ or equivalently $\epss(\gamma)$, applies also for the concentration and dielectric profiles.  We first present our results  for negative  $\beta$, which is the more prevalent case for most monovalent ions (e.g., halides and alkalines) (see Table~1).
\begin{figure*}
\includegraphics[width=0.82\textwidth]{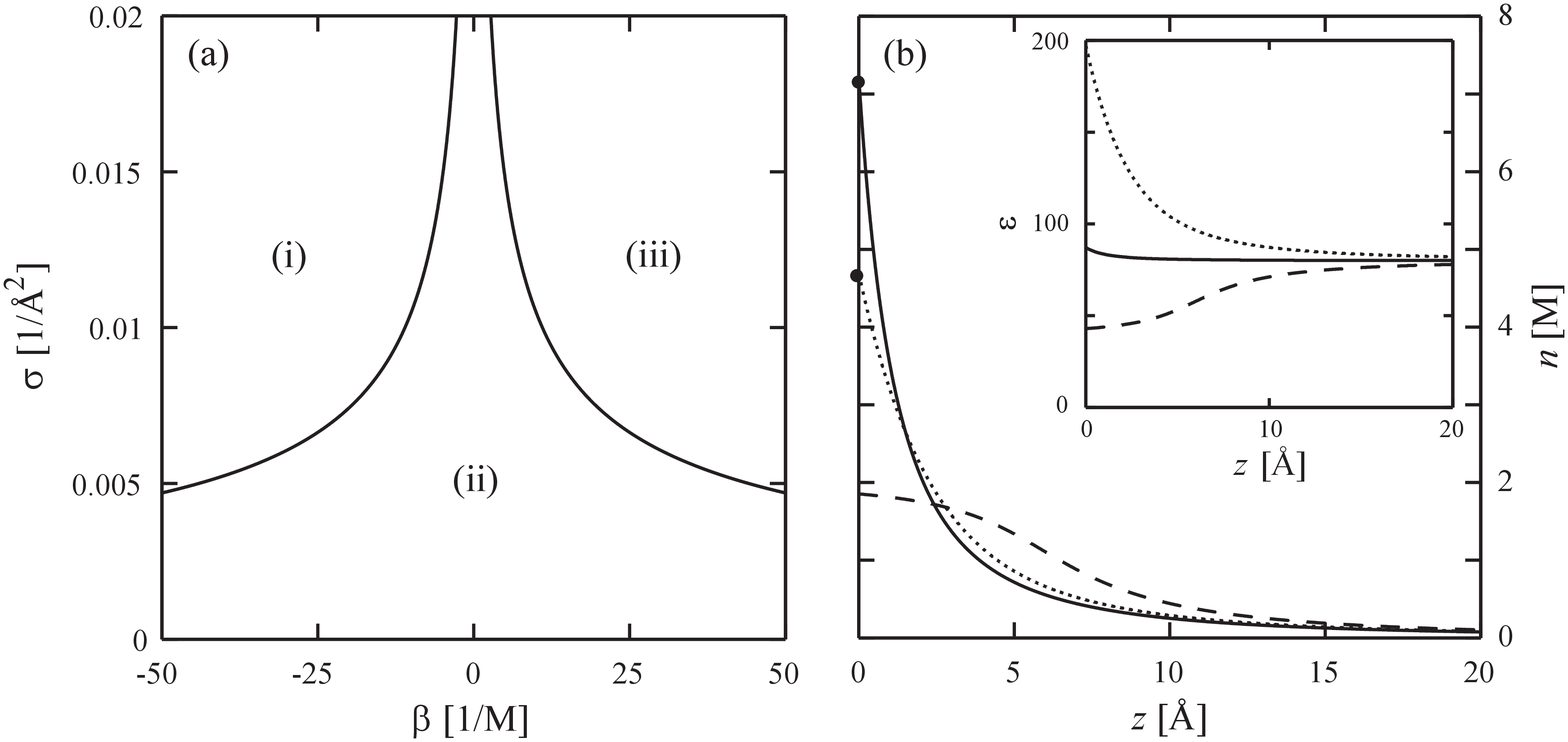}
\caption{\textsf{In (a) the regime diagram in the $(\beta,\sigma)$ plane in shown. Two cross-over lines separate the three limiting regimes: (\textit{i}), (\textit{ii}) and (\textit{iii}). In (b) typical concentration and dielectric constant profiles for each regime are shown, with surface charge density, $\sigma=0.01$\,\AA$^{-2}$. The dashed line corresponds to regime (\textit{i}) with $\beta=-20$\,M$^{-1}$, $n_s=1.9$~M and $\epss=43$. The solid line corresponds to regime (\textit{ii}) with $\beta=1$\,M$^{-1}$, $n_s=7.2$~M and $\epss=87$. The dotted line corresponds to regime (\textit{iii}) with $\beta=25$\,M$^{-1}$, $n_s=4.7$~M and $\epss=197$.}} \label{fig5}
\end{figure*}

The concentration profile is shown in Fig.~\ref{fig3}. It has a plateau in the vicinity of the surface, till about distances of $z\simeq5\,$\AA\ and larger, where there is an algebraic decay as in the regular PB profile (dashed line). The value of $\gamma=-1.8$ used in Fig.~\ref{fig3} is smaller than the crossover value, $\gamma_{12}=-1$ and lies within regime~({\it i}). For completeness we also present in Fig.~\ref{fig4} the $\beta>0$ case, where  $\gamma=2.3$ is larger than $\gamma_{23}=1$ [regime~({\it ii})]. Although the value of $n_s$ is reduced, the profile exhibits a regular algebraic decay, similarly to the PB profile (dashed line).

In the more common case of $\beta < 0$ (dielectric decrement by the ions), the ions should be depleted from the nearby wall region. This proximity behavior of the counter-ion cloud bears some resemblance to the case of steric effects~\cite{borukhov1997}, which leads to similar (but not identical) exclusion of counter-ions from the vicinal region.

The crossover values $\gamma_{12}$ can be transformed into relations between $\beta$ and the surface charge density $-\sigma$, yielding a regime diagram as shown in Fig.~\ref{fig5}(a). The crossover line between regimes (\textit{i}) and (\textit{ii}) is given by the relation:

\begin{equation}\label{sig12}
\sigma_{12}=\sqrt{\frac{\varepsilon_0}{2\pi\lb|\beta_{12}|}}\, ,
\end{equation}
which holds only for $\beta<0$. This crossover line separates between concentration profiles with a plateau-like behavior in the vicinity of the surface, to those where the concentration profile is nearly identical to the PB one. This crossover can be understood in terms of the dual role played by counter-ions as they accumulate at the surface. On one hand, just as in the standard PB model, the ions are attracted to the oppositely charged surface and diminish the local electrostatic field. On the other hand, for $\beta<0$ the ions  reduce the local dielectric constant and, hence, are repelled from the surface. In regime (\textit{i}), the latter prevails and leads to a plateau in the counter-ion density when $\sigma$ and $|\beta|$ are above the crossover line, as in Fig.~\ref{fig3}(a).

Similarly, the crossover line between regimes (\textit{ii}) and (\textit{iii}) is given by:
\begin{equation}
\sigma_{23}=\sqrt{\frac{\varepsilon_0}{2\pi\lb\beta_{23}}}\,,
\end{equation}
which holds only for $\beta_{23}>0$. In regime~({\it iii}) large values of $\beta$ increases substantially the value of $\varepsilon$ close to the surface [Fig.~\ref{fig4}(b)], but has no large effect on $n(z)\approx n_{\rm PB}(z)$. This somewhat surprising result is due to the attenuation of the electrostatic attraction for very large increase in the values of $\varepsilon$ at the surface.

\begin{figure*}[!tbh]
\includegraphics[width=0.4\textwidth]{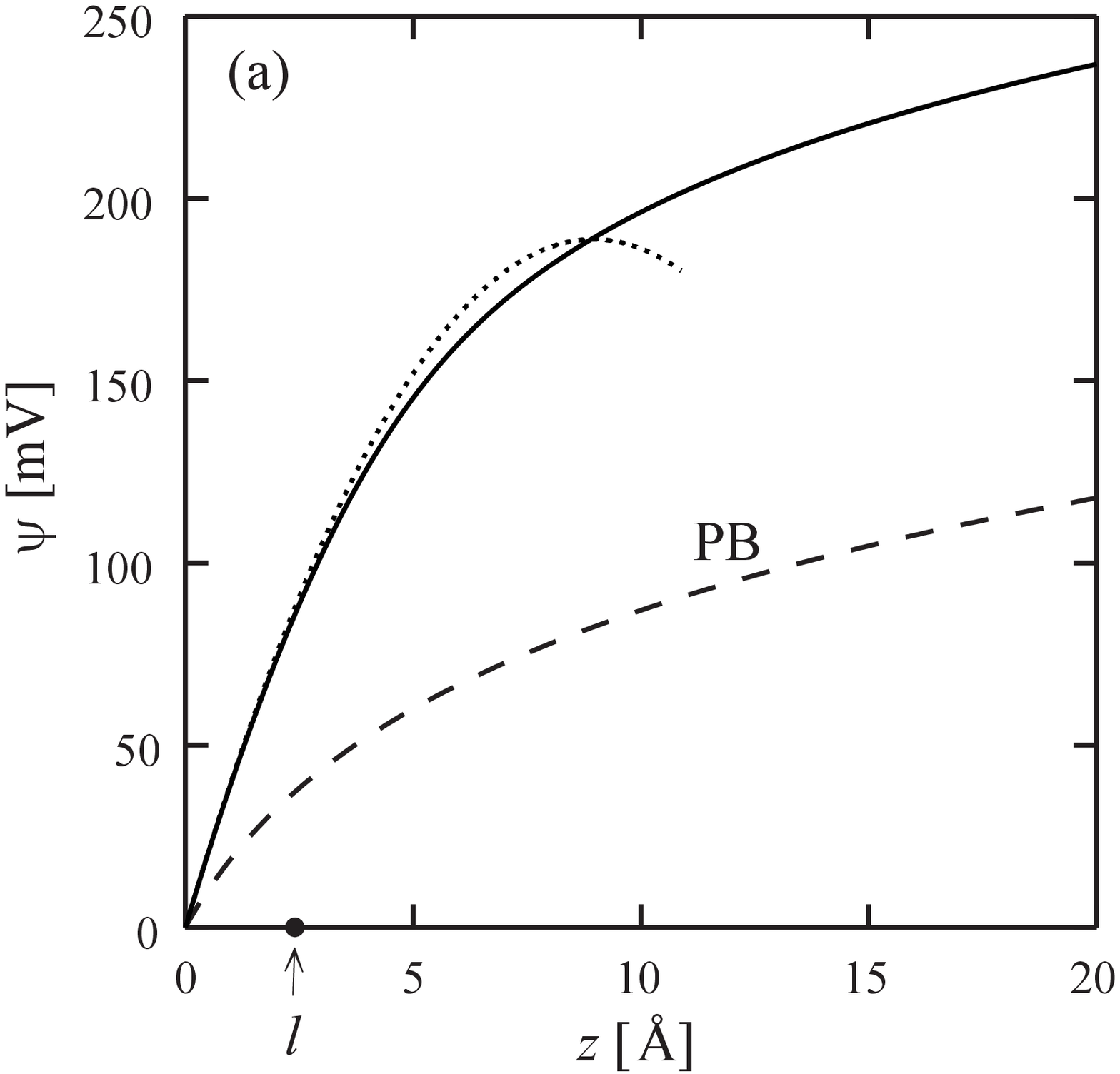}
\includegraphics[width=0.38\textwidth]{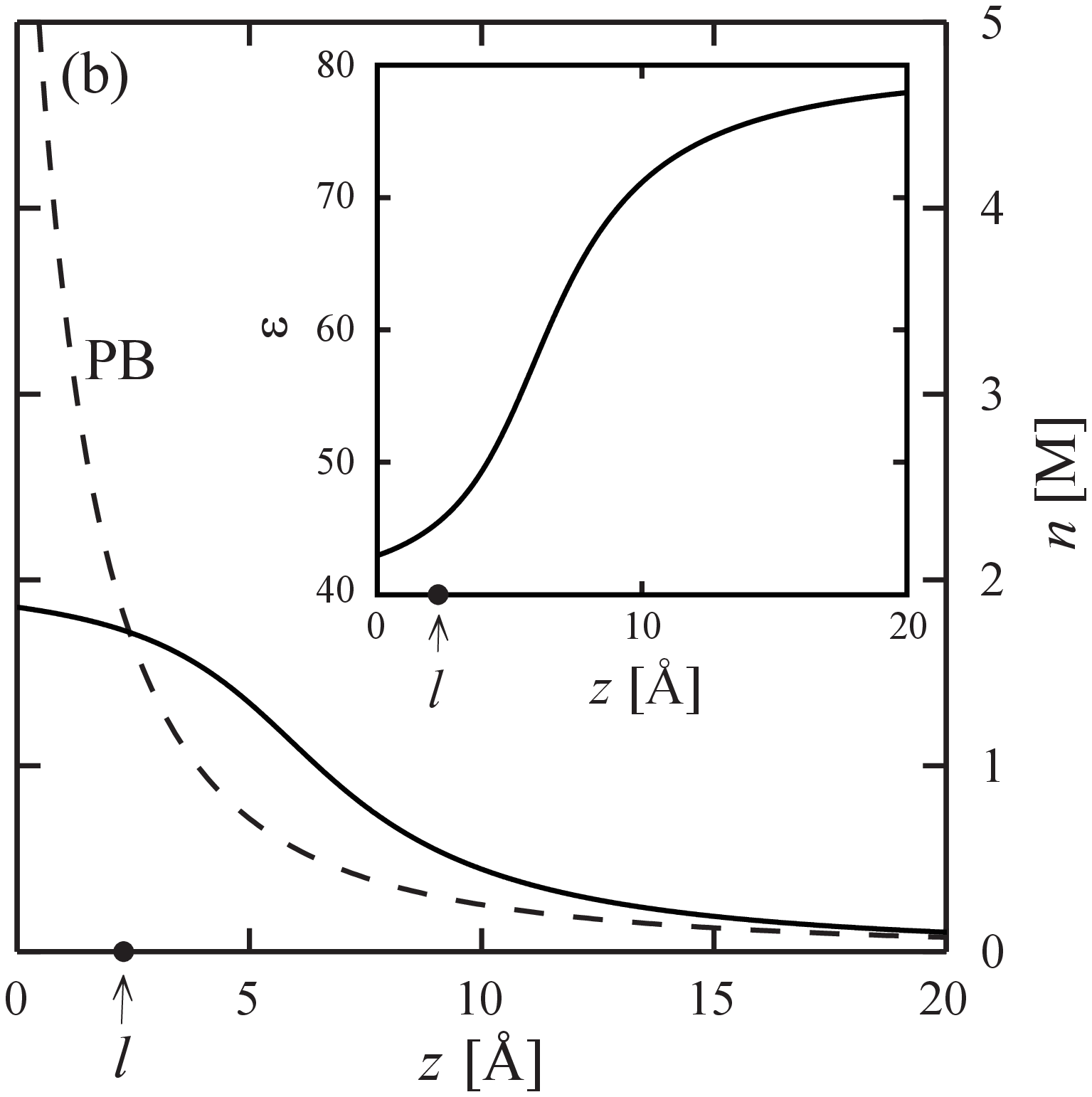}
\caption{\textsf{Profiles of electric potential, concentration and dielectric constant for parameter range as in regime (\textit{i}). In (a) the electrostatic potential $\psi$ is shown for $\sigma=0.01$\,\AA$^{-2}$ and $\beta=-20$\,M$^{-1}$. The solid line corresponds to the exact numerical solution, while the dotted line denotes the approximated one, Eq.~(\ref{approx_potential_regime1}), and the dashed line is the regular PB ($\beta=0$) result. In (b) and its inset, the concentration and dielectric constant profiles are shown,respectively, for the same parameters as in (a). The position of $l=2.3$\,\AA\ as calculated from
Eq.~(\ref{layer_thick}) is marked in (a), (b) and the inset.
It corresponds to the crossover from the slowly varying behavior to regular PB algebraic decay.}
} \label{fig6}
\end{figure*}

In Fig.~\ref{fig5}(b), we demonstrate the qualitative difference between the various regimes by plotting three corresponding counter-ion concentration profiles. For all shown profiles, the surface charge concentration is $\sigma=0.01$\,\AA$^{-2}$, while the parameter $\beta$ takes the values $\beta=-20,1$ and 25\,M$^{-1}$ for the dashed, solid and dotted lines, respectively.

For the profiles of regime (\textit{i}), $\beta<0$, one can calculate approximately the electrostatic potential in the vicinity of the charged surface by assuming that the counter-ion concentration varies slowly close to the surface and roughly obeys $n\approx n_s$. The potential is then given by the quadratic form:

\begin{equation}
\psi(z)\simeq \frac{2\pi e}{\varepsilon_0+\beta n_s}\left(-n_s z^2 +2\sigma z\right)\, .
\label{approx_potential_regime1}
\end{equation}
The width of the vicinal plateau, $l$, can  be estimated by noting that the electrostatic field for which the dielectric decrement becomes negligible is simply related to the crossover value, $\sigma_{12}$ given in Eq.~(\ref{sig12}). By demanding that $-\psi'(l)$, the  electrostatic field at $z=l$  matches $-4\pi e\sigma_{12}/(\varepsilon_0+\beta n_s)$, the plateau width $l$ is estimated to be:
\begin{equation}
\label{layer_thick}
l=2\sigma\frac{|\beta|}{\varepsilon_0}-\sqrt{\frac{2|\beta|}{\pi\lb\varepsilon_0}}\, ,
\end{equation}
where $n_s\simeq\varepsilon_0/(2|\beta|)$ in regime (\textit{i}), Eq.~(\ref{bc_1S_large_beta_minus}). The profile for $z>l\,$, can then be estimated as a regular PB profile with adjusted surface charge $\sigma=\sigma_{12}$ and a shifted $z$ axis: $z\rightarrow z-l$. The validity of Eqs.~(\ref{approx_potential_regime1}) and (\ref{layer_thick}) is examined in Fig.~\ref{fig6} by comparing them to the exact numerical results. Several observations are worth noticing. First, the value of $l\simeq 2.3$\,\AA\ is rather short-range. It gives an estimate to the width of the depleted ionic layer. Second, the assumption of a saturated ionic layer reproduces well the electric potential up to $z\simeq 3l$.

When compared to the corresponding PB results, the electrostatic potential shows a marked increase depending on the dielectric decrement. For parameters used in Fig.~\ref{fig6}, the electrostatic potential almost doubles in size as compared with its standard PB value. The effect persists for $z$ up to several dozens of Angstroms. Nevertheless, the more important effect of the electrostatic potential on the ionic density itself is much shorter ranged, and is attributed to the spatial variation of the dielectric response, see inset Fig.~\ref{fig6}(b).


\section{The pressure for the two-plate system}\label{pressure}

We now move on to analyze the case of two apposed charged planar surfaces. For simplicity sake, we  restrict ourselves to the symmetric case where two equally charged plates are located at $z=\pm D/2$, and evaluate the corresponding disjoining pressure $\hat{P}=P/\kbt$.

For an arbitrary function $\varepsilon(n)$ and $P\neq0$,
the boundary condition, Eq.~(\ref{bc}), is generalized as follows:
\begin{equation} \label{bc_general_2S}
n_{s}-\hat{P}=\frac{2\pi e^2 \sigma^2}{\kbt\varepsilon^2(n_{s})}
\left(\varepsilon(n_{s})+\frac{\D \varepsilon}{\D
n}\bigg|_{s} n_{s}\right)\,.
\end{equation}
This equation can then  be cast into a form that contains only $n_s$ and $\varepsilon(n_{s})$ without its derivatives~\cite{surface_integral}.
\begin{equation}
\frac{2\pi e^2 \sigma^2}{\kbt\varepsilon(n_{s}) n_{s}}
 +
\log{\frac{n_{s}}{\hat{P}}} + \frac{\hat{P}}{n_{s}}= {\rm const}.
\end{equation}
It shows that the pressure is in fact determined solely by the values of the surface energy
and the (ideal) entropy of mixing at the surface.

Employing the linearity assumption, $\varepsilon(n)=\varepsilon_0+\beta n$, the derivative of $f(n)$ is given by:
\begin{equation}
f'(n)=\frac{2 \beta^2
n(n-\hat{P})+(\epsz+\beta n)(\epsz+2\beta
n)}{2\sqrt{(n-\hat{P})(\epsz+2\beta n)^3}}\,
,\label{f_n_linear_2S}
\end{equation}
and the boundary condition, Eq.~(\ref{bc_general_2S}), can be expanded as a cubic polynomial in $n_s$:
%
\begin{eqnarray}
& \beta^2 n_s^3+(2\beta \varepsilon_0-\beta^2\hat{P})
n_s^2+\left(\varepsilon_0^2-2\beta\hat{P}\varepsilon_0-\frac{4\pi
e^2\sigma^2\beta}{\kbt}\right)n_s \nonumber \\
& -~\left(\hat{P}\varepsilon_0^2+\frac{2\pi
e^2 \sigma^2 \varepsilon_0}{\kbt}\right)=0\,
.\label{bc_general_linear_2S}
\end{eqnarray}

The counter-ion density profiles are given by Eqs.~(\ref{n_ode}), (\ref{f_n_linear_2S}) and (\ref{bc_general_linear_2S}), while the dependence of the pressure on the inter-plate separation $D$ is obtained by solving these equations for various values of $\hat{P}$, and inverting the function $D(\hat{P})$ into $\hat{P}(D)$.

Let us first verify that our theory has the correct PB limit for $\beta = 0$.
The differential equation of the density, Eq.~(\ref{n_ode}) reads

\begin{equation}
\label{rudi4}
\frac{\D n}{\D z}=-\sqrt{\frac{8\pi e^2}{\kbt \varepsilon_0}} n \sqrt{(n - \hat{P}_\mathrm{PB})}\, ,
\end{equation}
where  $\hat{P}_\mathrm{PB}$ is the standard PB pressure.
With $K= K(\hat{P}) = \sqrt{2\pi\lb\hat{P}}$, the solution of the above equation can be shown to be
\begin{equation}
\label{rudi4a}
\arctan{\sqrt{\frac{n(z)  - \hat{P} }{ \hat{P} }}} = - K z,
\end{equation}
or can be expressed as
\begin{equation}
\label{rudi5}
n(z) =  \hat{P} \left( 1 + {\tan^2 {K z}}\right) = \frac{ \hat{P} }{ {\cos^2{K z}}}.
\end{equation}
This is exactly the standard solution of the GC equation between two equally charged walls. The pressure is
then obtained by solving the boundary condition~\cite{andelman2006} that reduces to the following transcendental equation:


\begin{equation}
K D\tan \left(\frac{K D}{2}\right)=\frac{D}{\lgc}.
\end{equation}
%

\begin{figure}[!t]
\includegraphics[width=0.4\textwidth]{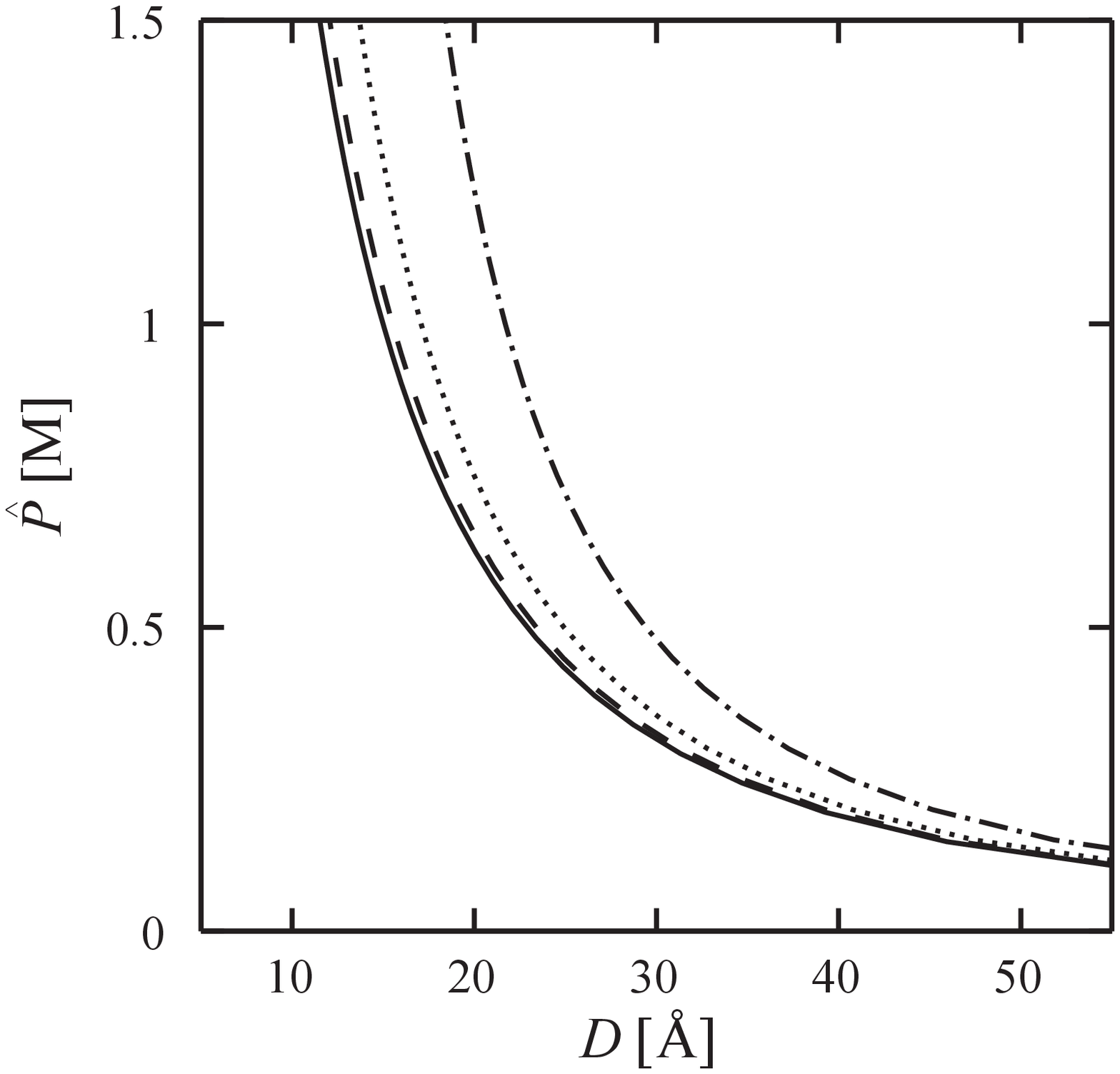}
\caption{\textsf{ The pressure $\hat{P}$ is shown as a function of the separation $D$ for surface charge density $\sigma=0.01$\,\AA$^{-2}$. The solid, dashed, dotted and dash-dotted
lines correspond to values of $\beta=0,-5,-10$ and $-20\,$M$^{-1}$, respectively. }} \label{fig7}
\end{figure}

For the general case of $\beta \neq 0$, the basic equation for the counter-ion density profile as well as the boundary condition have to be solved numerically. In Fig.~\ref{fig7} the dependence of the pressure $\hat{P}$ on the separation $D$ is shown for several values of $\beta$. The dashed line of $\beta=-5\,$M$^{-1}$ is almost identical to the standard PB prediction, given by the solid line ($\beta=0$). This implies that $\beta=-5\,$M$^{-1}$ is roughly the value where the dielectric decrement begins to affect the disjoining pressure. For more negative $\beta=-20\,$M$^{-1}$, the pressure strongly depends on the dielectric decrement and increases substantially with respect to the PB prediction. For example, at a separation $D=25\,$\AA, the pressure $\hat{P}=0.7\,$M with $\beta=-20\,$M$^{-1}$ is increased by 40\% comparing to $\hat{P}_{\mathrm{PB}}=0.42\,$M.

For small separations (and large pressure) the assumption of linear dependence of $\varepsilon$ on $n$ breaks down due to high values of $n >1.5$\,M. The results shown in Fig.~\ref{fig7} include only separations where the model is still valid. In the regime of extremely high surface counter-ion densities our model predicts unphysical results, such as negative local dielectric constant. This breakdown  does not imply any deep inconsistency but is just a straightforward consequence of the linearity {\sl ansatz}, Eq. (\ref{phenom}). This problem can easily be overcome by introducing a more general non-linear {\sl ansatz} with a saturation behavior.

\begin{figure*}[!t]
\includegraphics[width=0.8\textwidth]{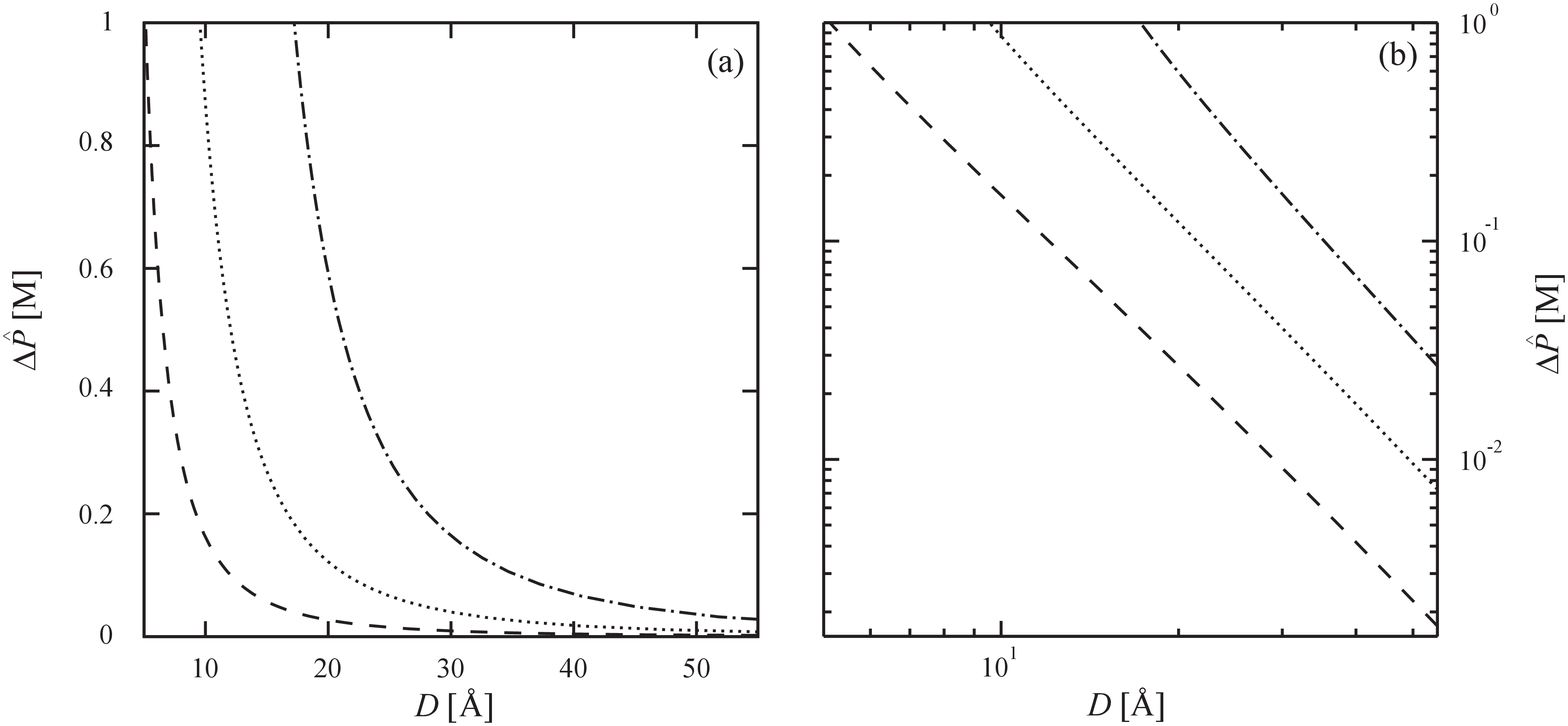}
\caption{\textsf{In (a) the  pressure deviation as compared to the PB pressure ($\beta=0$), $\Delta \hat{P}=\hat{P}-\hat{P}_\mathrm{PB}$, is plotted as function of $D$, for surface charge density $\sigma=0.01$\,\AA$^{-2}$.
In (b) similar results as in (a) are plotted on a log-log scale. The dashed, dotted and dash-dotted lines correspond to values of $\beta=-5,-10$ and $-20\,$M$^{-1}$, respectively.}} \label{fig8}
\end{figure*}

The deviation $\Delta\hat{P}=\hat{P}-\hat{P}_{\rm PB}$ of the pressure with respect to the PB pressure $\hat{P}_\mathrm{PB}$, is plotted in Fig.~\ref{fig8}(a) as a function of $D$. The $\Delta\hat{P}$ deviation  is significant for separations as large as a few nanometers. For $D<30$\AA\ the relative deviation may be as large as 65\%, while for $D\simeq50\,$\AA\ the deviation is smaller than 0.05\,M, leading to corrections smaller than 20\%.  In Fig.~\ref{fig8}(b), the deviation is plotted on a log-log scale, where $\Delta \hat{P}$ shows a power-law decay, $\Delta \hat{P}\sim D^\alpha$. With a linear fit the extracted $\alpha$ exponent  is $\alpha=-2.7,-2.8$ and $-3.1$ for $\beta=-5,-10$ and $-20\,M^{-1}$, respectively.

For $\beta\ll-1$ one can estimate the pressure analytically using  rescaled separation $D_{\mathrm{eff}}$ and
surface charge density $\sigma_{\mathrm{eff}}$. We consider an approximated profile with a plateau near the surface, having a thickness:
\begin{equation}
l=\frac{\sigma-\sigma_{12}}{n_s}\, ,
\end{equation}
employing the same arguments as in section \ref{sec_profile_beta}. Note that the values of $n_s$ and $\sigma_{12}$ are now given by Eq.~(\ref{bc_general_linear_2S}) instead of Eq.~(\ref{bc_1S}). The pressure associated with this profile can be calculated from the region where the behavior is PB-like, namely, for $-D/2+l<z<D/2-l$. This region can be considered as an independent profile with a surface charge density $\sigma_{\mathrm{eff}}=\sigma_{12}$ and separation $D_{\mathrm{eff}}=D-2l$. In analogy with the two limiting regimes of the standard PB theory \cite{andelman2006}, we find them here as well. First, ideal-gas regime where $D_{\mathrm{eff}}\ll{\lambda}^{\mathrm{eff}}_{\mathrm GC}$ and the pressure depends inversely on the separation: $\hat{P}\simeq\sigma_{\mathrm{eff}}/D_{\mathrm{eff}}$, where  the rescaled GC length is defined as ${\lambda}^{\mathrm{eff}}_{\mathrm GC}=1/(2\pi\lb\sigma_{\mathrm{eff}})$.
Second, when $D_{\mathrm{eff}}\gg{\lambda}^{\mathrm{eff}}_{\mathrm GC}$, the pressure dependence is given
by the standard Gouy-Chapman result valid for sufficiently large inter-surface separations, i.e., $\hat{P}\simeq \pi/(2\lb D_{\mathrm{eff}}^{\,2})$.


\begin{figure*}[!t]
\includegraphics[width=0.8\textwidth]{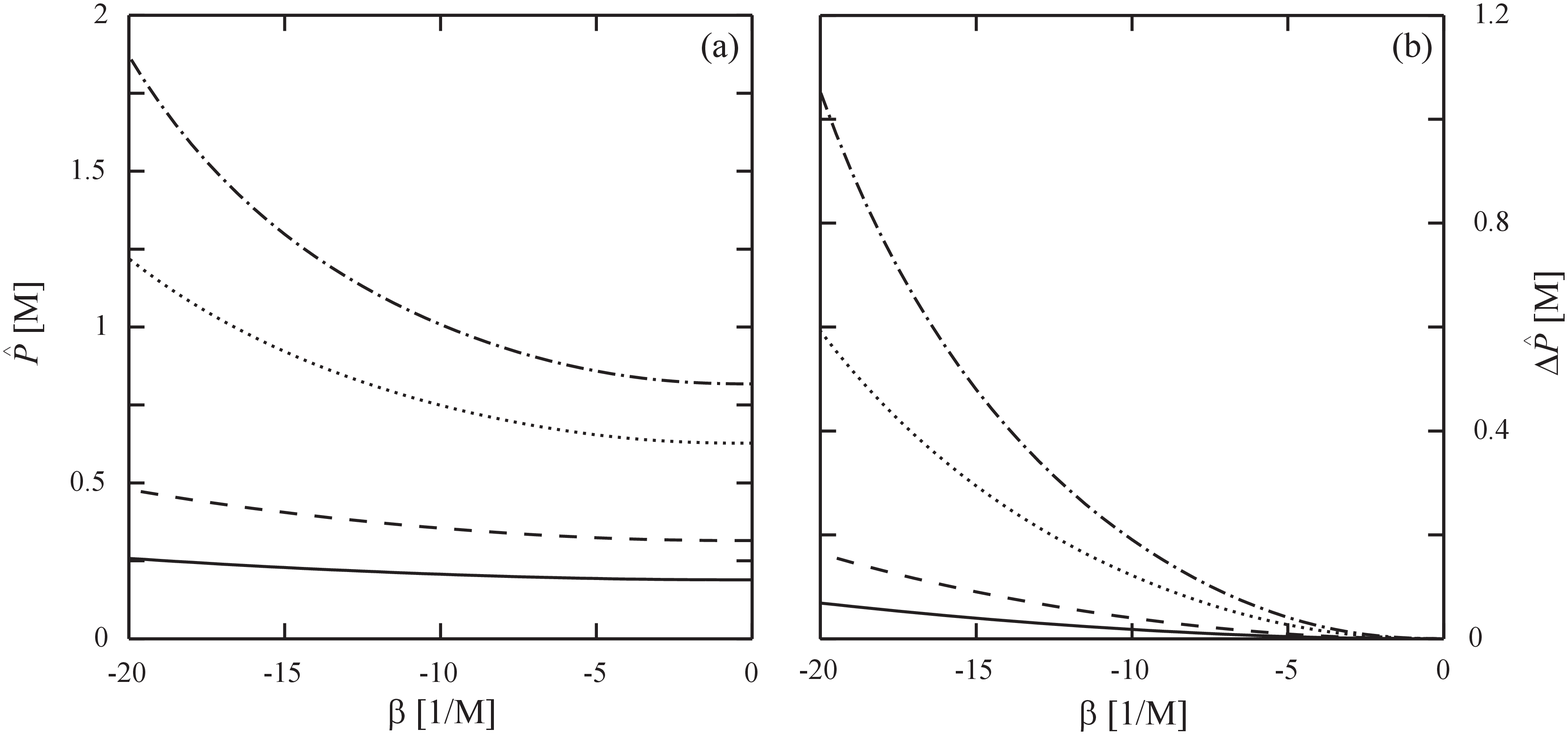}
\caption{\textsf{In (a) the total pressure $\hat{P}$ is shown as a function of the parameter $\beta$ for surface charge density $\sigma=0.01$\,\AA$^{-2}$. In (b) the pressure correction, $\Delta \hat{P}=\hat{P}-\hat{P}_\mathrm{PB}$, as compared to the PB pressure,  is similarly plotted. The solid, dashed, dotted and dash-dotted lines correspond to values of $D=40, 30, 20$ and $17\,$\AA, and standard PB pressure $\hat{P}_{\mathrm{PB}}=0.82, 0.63, 0.31$ and $0.19\,$M respectively.}} \label{fig9}
\end{figure*}

The pressure as a function of the parameter $\beta$ is presented in Fig.~\ref{fig9}(a) for several values of the separation $D$. The significance of the dielectric decrement effect becomes substantial for $\beta<-5\,$M$^{-1}$ as can be seen in the plot. For smaller values of $\beta$ the deviation from the standard PB value, $\hat{P}(\beta=0)$ is small and negligible for all the values of $D$. The magnitude of the deviation $\Delta\hat{P}$ as a function of $\beta$ is presented in Fig.~\ref{fig9}(b). For $\beta<-11.5\,$M$^{-1}$, the deviation is of the order of at least 10\% of the total pressure for all the values of $D$. For  example, for $D=30$\,\AA\ and $\beta=-11.5\,$M$^{-1}$, the pressure  $\hat{P}=0.37$\,M and the deviation is $\Delta\hat{P}=0.053$\,M.

\begin{figure}[!t]
\includegraphics[width=0.4\textwidth]{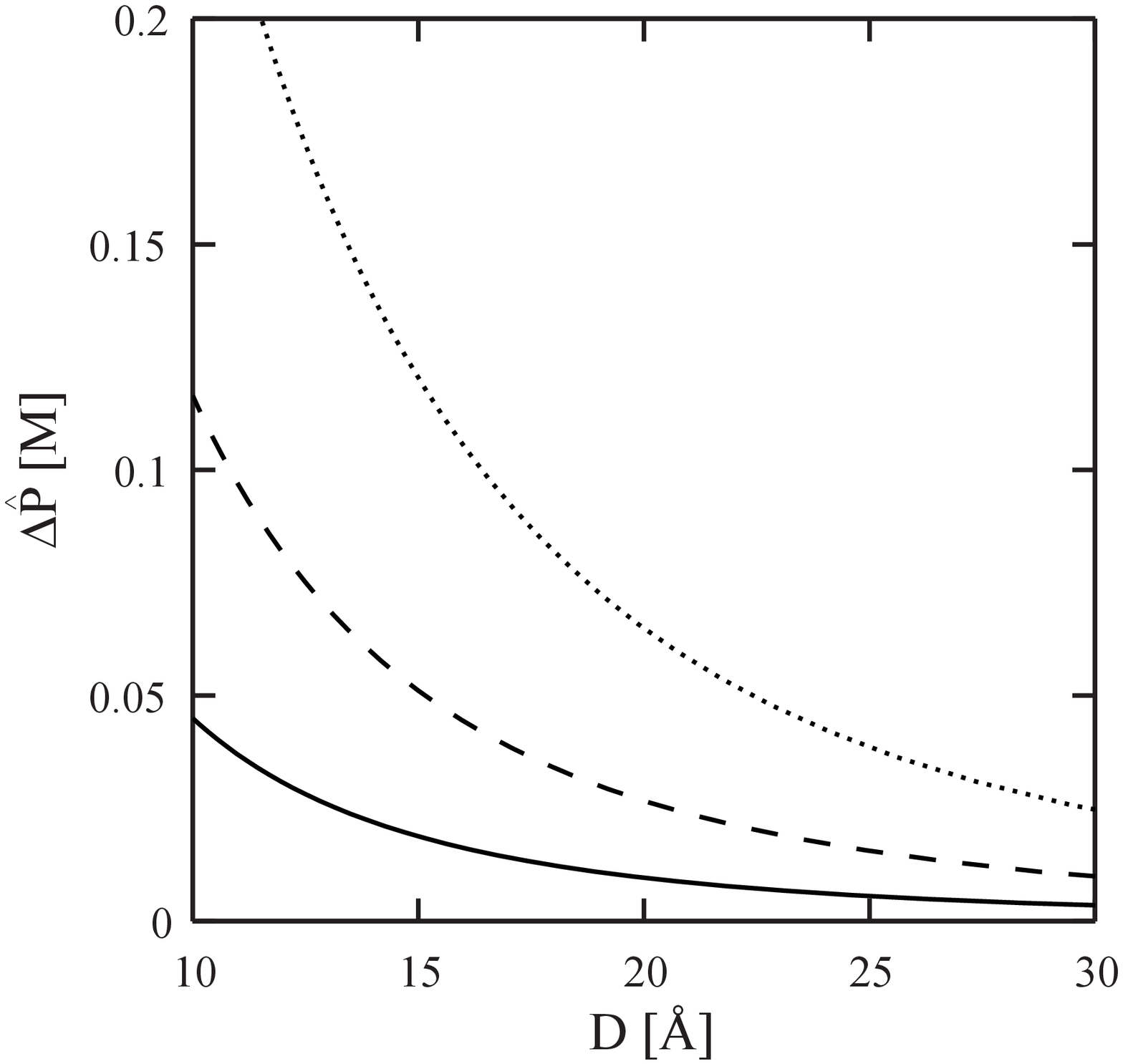}
\caption{\textsf{ The  pressure correction, $\Delta \hat{P}=\hat{P}-\hat{P}_\mathrm{PB}$ as a function of $D$,  is plotted for several $\beta>0$ values: solid, dashed and dotted lines correspond to $\beta=5, 10$, and 20.}} \label{fig10}
\end{figure}

For completeness we present  pressure profiles for $\beta>0$ in Fig.~\ref{fig10}. It is evident from comparing Fig.~\ref{fig8}(a) and Fig.~\ref{fig10} that for positive values of $\beta$ the deviation $\Delta \hat{P}$ is much smaller. This is in agreement with the analysis of the density profiles (see Sec.~IV.C), where a similar small deviation from the standard PB profile is found for $\beta>0$.

\section{Conclusions}

The model presented in this work accounts for  local changes in the dielectric constant of a solution due to the presence of ions with an effective polarization effect. Assuming linear dependence of the dielectric constant $\varepsilon$ on the ionic concentration $n$, we introduce a phenomenological parameter $\beta$ that describes the relation between $\varepsilon$ and $n$: $\varepsilon=\varepsilon_0+\beta n$. This parameter is ion-specific, and as shown in several experiments and computer simulations, its value is negative and varies between $-8$ and $-21$\,M$^{-1}$.

The concentration dependence of the dielectric constant leads to an additional coupling between the counter-ion concentration and the electrostatic potential. Consequently, the counter-ions play a dual role. On one hand their net charge induces an attractive interaction with the charged surface. On the other hand, the counter-ions effective dipole moment due to the concentration dependence of the dielectric constant leads to another electrostatic interaction that could be either attractive for $\beta>0$, or repulsive for $\beta<0$. The interplay between theses two interactions leads to changes of the free-energy minimal configuration.

Analyzing the behavior in ($\beta$,$\sigma$) parameter plane, we find three regimes. The most pronounced effect is found for regime (\textit{i}) and $\beta\ll-1$ , where ions are strongly depleted from the charged surface due to the decrement of $\varepsilon$ and the resulting penalty of electrostatic energy. This depletion results in a plateau of the ionic concentration profile close to the charged surface. We estimate the plateau thickness analytically and find that it is of the order of few angstroms. In regime (\textit{ii}) where $|\beta|<1$, the effect of local changes of the dielectric constant is negligible comparing to the contribution that comes from the net charge of the ions. The profiles in this regime are characterized by an algebraic decay similarly to the standard PB theory. For $\beta\gg 1$ in regime (\textit{iii}), the dielectric constant is increased nearby the surface, leading to a weaker attraction of the ions to the surface. The surface concentration is reduced comparing to the standard PB theory, but the overall change in the profiles is small.

Calculating the disjoining pressure between two planar surfaces, we find a substantial increase for negative values of $\beta$ of the order of $-10$. The deviation of the pressure calculated in our model as compared to the standard PB pressure reaches $\sim200\%$ for $\beta=-20$ and a separation $D\simeq18$\AA. The range of this additional effect persists up to a few nanometers. For $\beta=-10$\,M$^{-1}$, the deviation is significant for separations smaller than $\sim30$\AA. The dependence of the correction $\Delta\hat{P}$ on the separation $D$ exhibits an inverse power-law decay. The power varies between $-2.7$ to $-3.1$ for different values of $\beta$ as discussed in the previous section. For positive values of $\beta$ we find that the pressure does not vary substantially comparing to the standard PB prediction.

It is rather important to notice that the effect of the dielectric decrement on the disjoining pressure shows up as an effective ``solvent structural force" \cite{pars-water}, see Fig.~\ref{fig7}, between the two charged surfaces that acts at relatively small inter-plate separations, though no structural forces were assumed {\sl a priori} in our approach. This might signal a more general relationship between the dielectric decrement and water structure effects. In the present formulation, where the dielectric decrement itself does not contain any spatial scale, the dependence of the additional ``water structural force" on the inter-plate separation appears to scale with an inverse power of this separation.
%
%
However, for a model where the dielectric decrement levels off for larger values of the counter-ion density,  the additional phenomenological constant will introduce a new length scale into the problem. Then, possibly also the additional ``water structural force" would show such a length scale.

Further refinements and applications can be considered. For example, the model can be generalized to include salt ions, instead of  counter-ions only  as was done here. This will introduce another screening effect (the Debye length) that will compete with the other contributions.  Moreover, the concentration dependence of the dielectric constant can be taken to reproduce the real experimental results in order to improve the quantitative accuracy. Finally, by generalizing the model to curved geometries it will be possible to calculate  potential of mean force between spheres and cylinders.

The results presented in this work suggest that the effect of ions on the local dielectric constant should be taken into account on equal footing as other ion-specific interactions such as dispersion and hydration. Furthermore, this model may serve as a platform for a more detailed models, taking into account other ion-specific effects.

\bigskip
{\em Acknowledgement:~~~} We would like to thank Daniel Harries for many suggestions and comments that
were in particular instrumental in the initial stage of this work.
This work was partially supported by the U.S.-Israel Binational Science Foundation under Grant No.
2006/055, the Israel Science Foundation under Grant No. 231/08, and the
Agency for Research and Development of Slovenia (ARRS) through Program P1-0055
and Research Project J1-0908.

\appendix*
\section{The DFT formulation for two-plate system}\label{model_DFT}

We would like to show how to extend our formulation in Sec.~\ref{model} to the case where other interactions, besides electrostatic ones, also depend on the density profile, $n(z)$. Just as in Sec.~\ref{pressure}, the equally charged plates are taken to be at $z=\pm D/2$ yielding that $\psi'(z = 0) = 0$ at the symmetric midplane, $z =0$. The first Euler--Lagrange (EL) equation, Eq.~(\ref{eqlbrm_eq_1}), still remains valid

\begin{equation}
\psi'(z) = - \frac{4\pi e}{\varepsilon[n(z)]}\int_0^z n(z') dz'.
\label{field}
\end{equation}
A second integration would then yield $\psi(z)$ as a function of $n(z)$. Furthermore, it follows from the same mentioned EL equation that

\begin{equation}\label{DFT2}
e n \psi = - \frac{d}{dz}\left(  \frac{\varepsilon(n)}{4 \pi} \psi \psi'\right) + \frac{\varepsilon(n)}{4 \pi} \psi'^2.
\end{equation}
Instead of performing the minimization on the free energy with respect to the density field, $n$, as in Eq.~(\ref{eqlbrm_eq_2}), we will first express the free energy as a functional only of $n$. To be followed only then by a free energy variation. Inserting Eq.~(\ref{DFT2}) into the square brackets of the free energy, Eq.~(\ref{free_energy}),  the free energy becomes

\begin{eqnarray}
& F/{\cal A} ~=~ \int \D z\,\left[\frac{\varepsilon(n)}{8 \pi} \psi'^2 -
\frac{\D}{\D z}\left(  \frac{\varepsilon(n)}{4 \pi} \psi \psi'\right) \right. \nonumber \\
& \left. +~ k_BT n (\log{n} - 1) - \mu n\right] ~+~ \Sigma_s e \sigma \psi_s\, .
\label{interimfe}
\end{eqnarray}
%
By integrating explicitly the second term in Eq.~(\ref{interimfe}) it is easy to see that this contribution exactly cancels the surface term. The free energy per unit surface area then assumes the form
%
\begin{eqnarray}
{F}/{\cal A} &=&  \int \left( \frac{\varepsilon(n)}{8 \pi} \psi'^2
+~ k_BT n (\log{n} - 1) - \mu n\right) dz.\nonumber\\
& &
\end{eqnarray}
%
Substituting the expression for the local $\psi'$ field, Eq.~(\ref{field}), we get

\begin{eqnarray}
& {F}[n]/{\cal A} = \int \left[ \frac{4\pi e^2}{2 ~\varepsilon[n(z)]} \left(\int_0^z n(z') dz'\right)^2 \right.\nonumber\\
&\left. +~k_BT n(z) (\log{n(z)} - 1) - \mu n(z)\right] dz \, .
\end{eqnarray}
%
The free energy is now a functional of the local density $n(z)$ only. The variation of the above functional
with respect to $n(z)$ leads to

\begin{equation}
e\psi -\frac{1}{8\pi}\frac{\D \varepsilon(n)}{\D n}\psi'^2 +k_BT\log n - \mu = 0,
\end{equation}
which is exactly the second EL equation, Eq.~(\ref{eqlbrm_eq_2}), derived in Sec.~\ref{model} within the PB formulation.

The two formulations are indeed completely equivalent. The DFT formulation is preferable when one has additional terms in the free energy that depend either on the density or its derivatives, but this venue is left for future work.

\newpage




\begin{thebibliography}{99}
\bibitem{phenom}  In his book \cite{debye}, Debye discusses the dependence of the dielectric constant on the ionic concentation by treating the ions as cavities with a different value of dielectric constant compared to the surrounding solution. The results show that the value of $\beta$ dependes on the interior and exterior dielectric constant and on the size of the cavity. According to Debye, for most of the salts, the interior dielectric constant should be taken much smaller ($\varepsilon=1$) than the aqueous one ($\varepsilon=80$), leading to $\beta<0$. The delicate point is how to determine the size of the cavity, and Debye discusses that at length. However, the final conclusion of Debye is that there is no clear conclusion about this issue.
\bibitem{pos_beta}  The lowering of the surface density of ions is a consequence of two separate effects, which become apparent if the equilibrium density of ions is written in the form: \\
    (i) $n=n_0(\psi^{\prime})\exp(-e\psi/\kbt)$. One is the direct effect of the dielectric decrement, which leads to the changes of the Boltzmann distribution, as can be seen from: \\(ii) $\log n_0(\psi^{\prime})=\mu/\kbt+\beta(\psi^{\prime})^2/(8\pi \kbt)$. On the other hand, there is also an indirect effect of the dielectric decrement on the potential itself that can be obtained from the Poisson equation: (iii) $4\pi e n+(\varepsilon(n)\psi^{\prime})^{\prime}=0$. The total effect of the dielectric decrement-increment then depends on the combined action of both mechanisms and is obviously very non-linear. In fact one can see by combining (i) and (iii) that for non-vanishing $\beta$ the effect of dielectric decrement-increment is quadratic in $\beta$, and thus does not depend on its sign. See eq. (19) and regime (ii) in the adjacent text.
\bibitem{surface_integral}
Let us write the boundary condition in an alternative form. Noting first that
\\
%
$\varepsilon^{-2}(n_{s})[\varepsilon(n_{s})+\frac{\D \varepsilon}{\D
n}\big|_{s}\, n_{s}] = - n_{s}^2 \frac{d}{dn_{s}}[
1/(\varepsilon(n_{s}) n_{s})]
$
%
~and~
%
$
n_{s}=\hat{P} + n_{s}^2 \frac{d}{dn_{s}}[\log{(n_{s}/\hat{P})} + \hat{P}/n_{s}],
$
%
the equation for the surface counter-ion density $n_s$ can be written as transcendental relation that depends on the pressure $\hat{P}$ as a parameter:
%
$ 2\pi e^2 \sigma^2/(\kbt
\varepsilon(n_{s}) n_{s}) +
\log{(n_{s}/\hat{P})} + \hat{P}/n_{s}= {\rm const}.
$
%
Indeed, if multiplied by $n_{s}$, even in the case of a linear dependence of the dielectric function on the counter-ion concentration, the pressure can thus be written in a local form that contains only the surface electrostatic energy and the surface (ideal) entropy.
\end{thebibliography}
\end{document}